\newcommand{\dia}{\mbox{diag} }
\newcommand{\tr}{\mbox{Tr}\ }
\newcommand{\aaa}{A_\alpha}
\newcommand{\bi}{{\bar \psi}^i_t}
\newcommand{\bj}{{\bar \psi}^j_t}
\newcommand{\ha}{H_\alpha}
\newcommand{\gab}{g_{\alpha\beta}}
\newcommand{\gba}{g_{\beta\alpha}}
\newcommand{\la}{\Lambda_\alpha}

\newcommand{\jp}{\frac{1+(-1)^{N_t}}{2}}
\newcommand{\jm}{\frac{1-(-1)^{N_t}}{2}}
\newcommand{\be}{\begin{equation}}
\newcommand{\ee}{\end{equation}}
\newcommand{\calh}{{\cal H}}
\newcommand{\ra}{\rho_\alpha}
\newcommand{\ca}{{\cal A}}
\newcommand{\at}{{\cal A}_T}
\newcommand{\ats}{{\cal A}_{T*}}
\newcommand{\me}{{\cal M}(E)}
\newcommand{\hut}{\hat{U}_t}
\newcommand{\td}{\psi^2_t}
\newcommand{\ti}{\psi^i_t}
\newcommand{\tj}{\psi^j_t}
\newcommand{\tra}{\psi^1_t}
\newcommand{\trb}{{\bar \psi}^1_t}
\newcommand{\tdb}{{\bar \psi}^2_t}
\newcommand{\waoa}{W_{\alpha_0\alpha}}
\newcommand{\waob}{W_{\alpha_0\beta}}
\newcommand{\waoas}{\sum_{\alpha\neq\alpha_0}\waoa\waoa^{\, *}}
\newcommand{\half}{\frac{1}{2}}
\newcommand{\cpa}{{\bf C}P_\alpha}
\newcommand{\cpao}{{\bf C}P_{\alpha_0}}
\newcommand{\ebib}{e_{\beta,i_\beta}}
\newcommand{\ebjb}{e_{\beta,j_\beta}}

\documentstyle[12pt]{article}
\begin{document}
\begin{titlepage}
\today          \hfill
\begin{center}

\vskip .2in

{\large \bf On Uniqueness of the Jump Process in Event Enhanced Quantum
Theory}
\vskip .50in
\vskip .2in

A.~Jadczyk\footnote{
e-mail: ajad@ift.uni.wroc.pl}, G.~Kondrat and R.~Olkiewicz
\vskip .2in

{\em Institute of Theoretical Physics,
University of Wroc{\l}aw\\
Pl. Maxa Borna 9,
PL-50 204 Wroc{\l}aw}
\end{center}

\vskip .2in

\begin{abstract}
We prove that, contrary to the standard quantum theory of continuous
observation,
in the formalism of Event Enhanced Quantum Theory the stochastic process
generating sample histories of pairs (observed quantum system,observing
classical apparatus)is unique. This result gives a rigorous basis to
the previous heuristic argument of Blanchard and Jadczyk.
\end{abstract}
\end{titlepage}
\newpage
\section{Introduction}
Effective time evolution of a  quantum system is  usually described by
a dynamical semigroup: a semigroup of completely positive, unit preserving,
transformations acting on the algebra of observables of the system. A
general form of generator of a norm--continuous semigroup was published
in 1976 independently by Gorini, Kossakowski and Sudarshan \cite{koss}
(for matrix algebras) on one hand, and by Lindblad \cite{lin}
(for more general, norm-continuous case) on the other. It is usually
referred to as the Lindblad form; it reads:

\be
{\dot A} = i[H,A]+\sum_\alpha V_\alpha^\ast A V_\alpha - \frac{1}{2}
\{\Lambda , A\}
\ee
where $H=H^\ast$ is the Hamiltonian, $\{\, ,\, \}$ stands for anticommutator,
and
\be
\Lambda = \sum_\alpha V_\alpha^\ast V_\alpha.
\ee

In a contrast to a pure unitary evolution that describes closed systems
and which is time-reversible,
the second, dissipative part of the generator makes the evolution
of an open system irreversible. This irreversibility is not evident
from the very form of the equation, it
is connected with the positivity property of the evolution.
Formally we can often solve the evolution
equation backward in time, but positivity of the reversed evolution
will be lost. \\
We can also look at the dual time evolution of states rather than of
observables.
For states, described by density matrices, we get:
\be
{\dot \rho} = -i[H,A]+\sum_\alpha V_\alpha \rho V_\alpha^\ast - \frac{1}{2}
\{\rho , \Lambda\},
\ee
where the duality is defined by $\tr ({\dot A}\rho) = \tr ( A {\dot \rho})$.

Here again only propagation forward in time is possible, when we try
to propagate backward, then we will have to deal with negative
probabilities.
This irreversibility is reflected in the fact that pure states evolve into
mixed states.
How do mixed states arise? In quantum theory, similarly as in the
classical theory, they arise when we go from individual description to
ensemble description, from maximal available information to partial
information. Or simply, they arise by mixing of pure states.
Pure states are represented by one dimensional projection operators $P$.
If
$d\mu(P)$ is a probabilistic measure on pure states, then
the density matrix $\rho$ defined by $\rho=\int Pd\mu (P)$ is a mixed
state, unless $d\mu (P)$ is a Dirac
measure. Contrary to the classical theory, however, in quantum theory
decomposition of a mixed state into pure ones is non--unique. So,
for instance, the identity operator can be decomposed into {\em any}
complete orthonormal basis: $I=\sum_i \vert i><i\vert $, thus in
indenumerably many ways. This mysterious and annoying non-uniqueness of
decomposition into pure states in quantum theory can be simply taken as an
unavoidable price for our progress from classical to quantum, as a fact of
life. And so it was. Yet it started to cause problems in quantum measurement
theory.

The first attempt to give a precise mathematical formulation of quantum
measurement theory must be ascribed to John von Neumann.
In his monograph \cite{neu} he introduced two kind of evolutions: a
continuous, unitary evolution $U$ of an `unobserved' system, and
discontinuous `projections' that accompany `observations' or `measurements'.
His projection postulate, later reformulated by L\"uders for mixed states,
is expressed as follows:

`if we measure a property  $E$ of the quantum system, and if we do not make
any filtering which depends on the result, then as the result of this
measurement the system which was previously described by a density matrix
$\rho$ switches to the new state described by the density matrix
$E\rho E/\tr E\rho E$.'

A whole generation of physicists was brainwashed
by this apparently precise formulation. Few dared to ask: who are
`we' in the phrase `if we measure' \cite{gell94}, what is `measurement'
\cite{bell89,bell90}, at which particular
instant of time the reduction takes place? How long does it take \cite{paz},
if ever \cite{peres}, to reduce?  Can it be observed? Can it be verified
experimentally \cite{ballent,kartner,blaja95f}? Nobody could satisfactorily
answer these
questions. And so it was taken for granted that quantum theory can not really
be
understood in physical terms, that it is a peculiar mixture of objective
and subjective. That it is about `observations', and so it makes little
or no sense without `observers', and without `mind'. There were many that
started to believe that it is the sign of new  age and the sign of progress.
Few opponents did not believe completeness of a physical theory that could
not
even define what constitutes `observation' \cite{bell89,bell90}--
but they could not change the overall feeling of satisfaction with successes
of
the quantum theory.

This situation started to change rapidly when technological progress make it
possible to make prolonged experiments with individual quantum systems.
The standard `interpretation' did not suffice. Experimenters were
seeing with their own eyes not the `averages' but individual sample
histories. In particular, experiments in Quantum Optics allowed one to
almost `see' the quantum jumps. In 1988 J.R. Cook \cite{cook88} discussed
photon counting statistics in fluorescence experiments and revived
the question `what are quantum jumps?'. Another reason to pay
more attention to the notion of quantum jumps came from the several
groups of physicists working on effective numerical solutions of
quantum optics master equations. The works of Carmichael \cite{carm93},
Dalibard, Castin and M{\o}lmer
\cite{dal92,molmer93}, Dum, Zoller and Ritsch \cite{dum92},
Gardiner, Parkins and Zoller \cite{gard92},
developed the method of Quantum Trajectories, or Quantum Monte Carlo (QMC)
algorithm
for simulating solutions of master equations.
It was soon realized
(cf. e.g. \cite{gisin84,gisin89,diosi88,diosi89,pearle89})
that the same master
equations can be simulated either by Quantum Monte Carlo method
based on quantum jumps, or by a continuous quantum state diffusion.
Wiseman and Milburn \cite{wise93}
discussed the question of how different experimental detection schemes
relate  to continuous diffusions or to
discontinuous jump simulations. The two approaches were recently
put into comparison also by Garraway and Knight \cite{garr94}. There
are at present two schools of simulations. Gisin et al. \cite{gisin93}
tried to reconcile the two arguing that \lq {\em the quantum jumps
can be clearly seen}\rq\/ also in the quantum state diffusion plots.
On the other hand already in 1986 Diosi \cite{dio1}
proposed a pure state, piecewise deterministic process that reproduces
a given master equation. In spite of the title of his paper that suggests
uniqueness of his scheme, his process although mathematically canonical for a
given master equation - it is not unique.\\
This problem of non-uniqueness is especially important in theories
of gravity-induced spontaneous localization (see \cite{ghira90}, also
\cite{pearle95a,pearle95b} and references therein) and in the recent
attempts to merge mind-brain science with quantum theory
\cite{stapp95,ham96,nanop95}, where quantum collaps plays an important role.

In the next section we shall see how the situation changes completely with
the new approach to quantum measurement developed by Ph. Blanchard and
one of us (see \cite{blaja95a} and references there)\footnote{
Complete, actual bibliography of the Quantum Future Project
is always available under URL: http://www.ift.uni.wroc.pl/~$\tilde{}$
ajad/qf-pub.htm}.
In Sec. 2 we will sketch the main idea of the new approach. We
will also indicate infinitesimal proof of uniqueness of the stochastic
process that reproduces master equation for the total system, i.e. quantum
system+classical apparatus. In Sec. 3 we give concrete examples of
non-unicity when only a pure quantum system is involved - as it is
typical in quantum optics. In Sec. 4. we will give a rigorous, global
proof of unicity of the process, when classical apparatus is coupled
in an appropriate way to the quantum system. Conclusions will be given in
Sec. 5. There we also comment upon the most natural question: we all
know that every apparatus consists of atoms - then how can it be
classical?
\section{The formalism}
Let us sketch the mathematical framework of the ``event-enhanced quantum
theory''. Details can be found in \cite{blaja95a}.
To describe events, one needs a classical system $C$, then possible events
are  identified with changes of a (pure) state of $C$. One can think
of events as `clicks' of a particle counter, changes of the pointer position,
changing readings on an apparatus LCD display. The concept of an event is
of course an idealization - like all concepts in a physical theory. Let us
consider the simplest situation corresponding to a finite set of possible
events. The space of pure states of $C$, denoted by ${\cal S}_c$, has $m$
states, labeled by $\alpha=1,\ldots,m$.
Statistical states of $C$ are probability measures on ${\cal S}_c$ -- in
our case just sequences $p_\alpha\geq 0, \sum_\alpha p_\alpha=1$. \\
The algebra of observables of $C$ is the algebra $\ca_c$ of complex
functions on ${\cal S}_c$ -- in our case
just sequences $f_\alpha, \alpha =1,\ldots,m$ of complex numbers.\\ We
use Hilbert space language even for the description of the
classical system. Thus we introduce an $m$-dimensional Hilbert space
$\calh_c$ with a fixed basis, and we realize $\ca_c$ as the algebra of
diagonal matrices $F=\dia(f_1,\ldots,f_m)$.\\ Statistical states of $C$ are
then diagonal density matrices $\dia(p_1,\ldots,p_m)$, and pure states of
$C$ are vectors of the fixed basis of $\calh_c$.\\ Events are ordered
pairs of pure states $\alpha\rightarrow\beta$, $\alpha\neq\beta$. Each
event can thus be represented by an $m\times m$ matrix with $1$ at the
$(\alpha,\beta)$ entry, zero otherwise. There are $m^2-m$ possible events.\\

\noindent We now come to the quantum system.\\
Let $Q$ be the quantum system whose
bounded observables are from the algebra
$\ca_q$ of bounded operators on a Hilbert space $\calh_q$.
In this paper we will assume  $\calh_q$ to be {\em finite dimensional}\ .
Pure states of $Q$ are unit vectors in $\calh_q$;
proportional vectors describe the same quantum state. Statistical
states of $Q$ are given by non--negative density matrices ${\hat\rho}$,  with
$\tr ({\hat\rho})=1$.

Let us now consider the total system $T=Q\times C$.
For the algebra $\ca_t$ of observables of $T$
we take the tensor product of algebras of observables of $Q$ and $C$:
$\ca_t=\ca_q\otimes\ca_c$. It acts on the tensor product
$\calh_q\otimes\calh_c=\oplus_{\alpha=1}^m\calh_\alpha$, where
$\calh_\alpha\approx\calh_q.$ Thus
$\ca_t$ can be thought of as algebra of {\em diagonal} $m\times m$
matrices $A=(a_{\alpha\beta})$, whose entries are quantum operators:
$a_{\alpha\alpha}\in \ca_q$, $a_{\alpha\beta}=0$ for $\alpha\neq\beta$.\\
Statistical states of $Q\times C$ are given by $m\times m$ diagonal matrices
$\rho=\dia(\rho_1,\ldots,\rho_m)$ whose entries are positive operators
on $\calh_q$, with the normalization $\tr (\rho)=\sum_\alpha\tr
(\ra)=1$.
Duality between observables and states is provided
by the expectation value $<A>_\rho=\sum_\alpha \tr (\aaa\ra)$.

We will now generalize slightly our framework. Indeed, there is no need
for the quantum Hilbert spaces $\calh_\alpha$, corresponding to different
states of the classical system, to coincide. We will allow them to
be different in the rest of this paper. We denote
$n_\alpha=dim(\calh\alpha).$

We consider now dynamics. It is normal in quantum theory that classical
parameters enters quantum Hamiltonian. Thus we assume that
quantum dynamics, when no information is
transferred from $Q$ to $C$, is described by Hamiltonians
$\ha :\calh_\alpha \longrightarrow \calh_\alpha$,
that may depend on the actual state of $C$ (as indicated by the index
$\alpha$).
We will use matrix notation and write $H=\dia(\ha)$.
Now take the classical system. It is discrete here.
Thus it can not have continuous time
dynamics of its own.

As in \cite{blaja95a} the {\em coupling} of $Q$ to $C$ is specified by a
matrix $V=(\gab)$, where $\gab$ are linear operators:
$\gab :\calh_\beta \longrightarrow \calh_\alpha$.
We put  $g_{\alpha\alpha}=0$. This condition expresses the simple
fact: we do not need dissipation without receiving information (i.e
without an event).
To transfer  information
from $Q$ to $C$ we need a non--Hamiltonian term which provides
a completely positive (CP) coupling. As in \cite{blaja95a} we consider
couplings for which the evolution
equation for observables and for states is given by the Lindblad  form:
\be
{\dot A}_\alpha=i[\ha,\aaa]+\sum_\beta \gba^\star
A_\beta \gba - {1\over2}\{\la,\aaa\},\label{eq:lioua}
\ee
or equivalently:
\be
{\dot \rho}_\alpha=-i[\ha,\ra]+\sum_\beta \gab
\rho_\beta \gab^\star - {1\over2}\{\la,\ra\},\label{eq:liour}
\ee
where
\be
\la=\sum_\beta \gba^\star \gba.
\ee

The above equations describe statistical behavior of ensembles. Individual
sample histories are described by a Markov process with values in pure
states of the total system. In \cite{blaja95a} this process was argued
to be infinitesimally unique. For the sake of completeness we repeat here
the arguments.
First, we  use Eq. (\ref{eq:liour}) to
compute $\ra (dt)$ when the initial state $\ra (0)$ is pure:
\be
\ra(0)=\delta_{\alpha\alpha_0}|\psi_0><\psi_0|.
\ee
 In the equations
below we will discard terms that are higher than linear order in $dt$.
For $\alpha=\alpha_0$ we obtain:
\be
\begin{array}{lrl}
\rho_{\alpha_0}(dt)=&
|\psi_0><\psi_0|&-i[H_{\alpha_0},|\psi_0><\psi_0|]\, dt- \\
\null&\null&-{1\over2}\{\Lambda_{\alpha_0},|\psi_0><\psi_0|\} \, dt\/,
\end{array}
\ee
while for $\alpha\neq\alpha_0$
\be
\rho_{\alpha_0}(dt)=
g_{\alpha\alpha_0}|\psi_0><\psi_0|g_{\alpha\alpha_0}^\star\, dt
\ee
The term for $\alpha=\alpha_0$ can be written as
\be
\rho_{\alpha_0}(dt)=p_{\alpha_0}|\psi_{\alpha_0}><\psi{\alpha_0}|,
\label{eq:a0}
\ee
where
\be
\psi_{\alpha_0}={
{\exp\left(-iH_{\alpha_0}dt-
{1\over2}\Lambda_{\alpha_0}dt\right)\psi_0}
\over{\Vert\exp\left(-iH_{\alpha_0}dt-
{1\over2}\Lambda_{\alpha_0}dt\right)\psi_0 \Vert}},
\ee
and
\be
p_{\alpha_0}=1-\lambda(\psi_0,\alpha_0 )dt.
\ee
The term with $\alpha\neq\alpha_0$ can be written as:
\be
\ra (dt)=p_\alpha \, |\psi_\alpha><\psi_\alpha|\, ,\label{eq:a}
\ee
where
\be
p_\alpha=\Vert g_{\alpha\alpha_0}\psi_0\Vert^2 dt,
\ee
and
\be
\psi_\alpha={
{g_{\alpha\alpha_0}\psi_0}\over{\Vert g_{\alpha\alpha_0}\psi_0\Vert}}
\ee
This representation is unique and it defines the infinitesimal version of
 a piecewise deterministic Markov process.
\section{Non-uniqueness in the pure quantum case}
In this section we will show on simple examples the nature of
non-uniqueness in the pure quantum case.

For simplicity let us consider a two state quantum system whose algebra
of observables is equal to $M_{2\times 2}$. Let $T_t$ be a dynamical
semigroup
with a generator $L$ given by
$$L(\rho )\;=\;a\rho a^{\ast}\;-\;\frac{1}{2}\{ a^{\ast}a\:,\:\rho\},$$
where $a\in M_{2\times 2}$.
\subsection{Pure diffusion process}
First let us show that the time evolution
determined by $L$ can be described by a diffusion process with values in
${\bf C}P^1$ \cite{gisin}.\\
Let a two component complex valued process $\psi_t\:=\:(\psi^1_t,
\psi^2_t)\prime$ (prime denotes the transposition) be given by the following
stochastic differential equation:
$$d\ti\; =\; f_i(\psi_t)dB_t\; +\; g_i(\psi_t)dt,\; \; i=1,2,$$
where $B_t$ is a one-dimensional real Brownian motion and
$$g_i(\psi_t)\:=\:\sum\limits_j
(<a^{\ast}>_ta_{ij}-\frac{1}{2}(a^{\ast}a)_{ij})\tj
\:-\:\frac{1}{2}<a^{\ast}>_t<a>_t\ti$$
$$f_i(\psi_t)\:=\:\sum\limits_j a_{ij}\tj\:-\:<a>_t\ti$$
$$<a>_t\:=\:\frac{<\psi_t|a|\psi_t>}{<\psi_t|\psi_t>},\quad
<a^{\ast}>_t\:=\:\frac{\psi_t|a^{\ast}|\psi_t>}{<\psi_t|\psi_t>}$$
Moreover let us choose an initial condition $\psi_0=(z^1_0,z^2_0)\prime$
such that $|z^1_0|^2+|z^2_0|^2=1$. Because $f_i$ and $g_i$ are continuously
differentiable (in the real sense) on ${\bf C}^2\setminus \{0\}$ so there
exists a local solution with a random explosion time T (see for example
\cite{protter}). But
$$d|\ti|^2\:=\:\ti d\bi\:+\:\bi d\ti\:+\:d[\ti,\bi]_t,$$
where $[\ti,\bi]_t$ is the quadratic covariation of $\ti$ and $\bi$. Thus
$$d[\ti,\bi]_t\:=\:|f_i(\psi_t)|^2dt,$$
and so $$d\|\psi_t\|^2\:=\:\sum\limits_i
(\ti d\bi\:+\:\bi d\ti)\;+\;\|f(\psi_t)\|^2dt\:=\:0.$$
It implies that $T=\infty$ with probability one and so our process is a
diffusion with values in a sphere $S^3$. Let us define a process $P_t$
with values in one dimensional projectors by
$$P_t=\:|\psi_t><\psi_t|\:=\:\sum\limits_{i,j}\ti\bj e_{ij},$$
where
$e_{ij}$ form the standard basis in $M_{2\times 2}$. Then, using the equation
$$d(\ti\bj)\:=\:({\bar f}_j\ti\:+\:f_i\bj)dB_t\;+\;({\bar g}_j\ti\:+\:
g_i\bj\:+\:f_i{\bar f}_j)dt$$
we obtain that $$dP_t\:=\:[(a-<a>_t)P_t\:+\:P_t(a^{\ast}-<a^{\ast}>_t)]dB_t
\:-\:\frac{1}{2}\{a^{\ast}a,\,P_t\}dt\:+\:aP_ta^{\ast}dt.$$
Since $B_t$ is a martingale then after taking the average we get
$$dE[P_t]\:=\:aE[P_t]a^{\ast}dt\:-\:\frac{1}{2}\{a^{\ast}a,\,E[P_t]\}dt$$
Let us define a density matrix $\rho_t=\:E[P_t]$. Then
$$\dot{\rho}_t\:=\:a\rho_ta^{\ast}\:-\:\frac{1}{2}\{a^{\ast}a,\,\rho_t\},$$
and so the average of the diffusion gives the quantum dynamical evolution.

Finally, we show that $\rho_t=\int P(t,\,x_0,\,dy)P_y$,
where $P_y=|y><y|$, $x_0=\:|\psi_0>
<\psi_0|$ and $P(t,\,x,\,dy)$ is the transition probability of the described
diffusion.
By the definition, $P(t,\,x_0,\,\Gamma)$ is the distribution of the
random variable $P_t^{x_0}$ such that $P_0^{x_0}=\:x_0$. It implies that for
every bounded and measurable function $f$ defined on ${\bf C}P^1$ we have
$$E[f(P_t^{x_0})]\:=\:\int f(y)P(t,\,x_0,\,dy).$$
Let us consider a function given by $f(y)=\:Tr(AP_y)$, where $A\in M_{2\times
2}$. Then $$\int Tr(AP_y)P(t,\,x_0,\,dy)\:=\:E[Tr(AP_t^{x_0})]\:=\:Tr(A
\rho_t).$$ So $Tr(A\int P(t,\,x_0,\,dy)P_y))\:=\:Tr(A\rho_t)$ for every $A$
and thus $\rho_t=\int P(t,\,x_0,\,dy)P_y$ with $\rho_0=\:x_0$.\\
\subsection{Piecewise deterministic solution}

On the other hand it is possible to associate with the same quantum dynamics
a piecewise deterministic process, as in the method of quantum trajectories
\cite{carm93}.  Now the situation is more complicated,
because, in general, we can not replace the Brownian motion by the Poisson
process. We have to solve a stochastic differential equation for an unknown
process $({\tilde N}_t,\:\psi_t)$.
$$d\ti\:=\:f_i(\psi_{t^-})d{\tilde N}_t\:+\:g_i(\psi_t)dt,$$
where $f_i$ and $g_i$ are prescribed functions, together with the following
constrain: ${\tilde N}_t$ is a semimartingale such that

a) $[{\tilde N},\,{\tilde N}]_t\:=\:{\tilde N}_t$, ${\tilde N}_0=\:0$,
$E[{\tilde N}_t]<\infty$ for all $t\geq 0$,\\

b) for a given nonnegative function $\lambda :{\bf C}^2\to {\bf R}$ the
process\\ $M_t:=\:{\tilde N}_t\:-\:\int^t_0 \lambda(\psi_s)ds$
is a martingale.

It is clear that $M_t$ will be a purely discontinuous martingale. A
continuous, increasing and with paths of finite variation on compacts
process: $\int_0^t\lambda (\psi_t)ds$ is called the compensator of ${\tilde
N}_t$. In our case due to assumption a) it is also the conditional quadratic
variation of ${\tilde N}_t$ \cite{protter}.
The functional $\lambda(\psi_t)$ is called the stochastic intensity and
plays the role of the intensity of jumps. Let
us recall that for the (homogeneous) Poisson process
$N_t\:-\int^t_0 \lambda ds\:=\:N_t\:-
\:\lambda t$ is a martingale. From the assumption a) above we obtain that
${\tilde N}_t$ is quadratic pure jump, its continuous part is equal zero and
$\triangle{\tilde N}_s\:=\:(\triangle{\tilde N}_s)^2$,
where $\triangle{\tilde
N}_s\:=\:{\tilde N}_s\:-\:{\tilde N}_{s^{-}}$ so it is a point process. Let
us emphasize that in general it is not an inhomogeneous Poisson process
since its compensator would be a deterministic function equal to $E[{\tilde
N}_t]$ \cite{jacod}. So it will be the case only when the stochastic
intensity is a deterministic function depending on $t$.\\
Moreover $[{\tilde N},\,t]_t\:
=\:0$ as ${\tilde N}_t$ is of finite variation on compacts. It implies the
following symbolic rules
$$(d{\tilde N})^2\:=\:d{\tilde N},\quad
d{\tilde N}dt\:=\:dtd{\tilde N}\:=\:0.$$
{}From assumption b) we get $dM_t=d{\tilde N}_t-\lambda (\psi_t)dt$. Let
${\cal F}_t$ be a $\sigma$-algebra of all events up to time $t$. Because
$M_t$ is a martingale, so $E[dM_t\vert{\cal F}_t]=0$ what implies:
$$E[d{\tilde N}_t\vert{\cal F}_t]=\lambda(\psi_t)dt$$
see \cite{barbel91}.

Till now the operator  $a\in M_{2\times 2}$ was arbitrary. A particular
simple case is if we take
$$a^{\ast}\:=\:a\:=\left(\matrix{0&1\cr 1&0\cr}\right)$$. Then
$L(\rho_t)\:=\:a\rho_ta\:-\:\rho_t$ and so the intensity
$$\lambda(\psi_t)\;=\;<a^{\ast}a>_t\;=
\;\frac{<\psi_t|a^{\ast}a|\psi_t>}{<\psi_t|\psi_t>}\;=\;1$$
what implies that ${\tilde N}_t\:=\:N_t$.
Because there is no deterministic evolution (we do not have the
Hamiltonian part and the jump rate is constant) so in this case we
can put $g_1\:=\:g_2\:=\:0$ and $f_1(\psi_t)\:=\:\td -\tra$,
$f_2(\psi_t)\:=\:
\tra -\td$ as the probability of a particular jump depends on the difference
between $\tra$ and $\td$. Thus we arrive at
$$d\ti\:=\:f_i(\psi_{t^-})dN_t.$$
Using the identity $d[\psi^i,\,\psi^j]_t\:=\:f_i{\bar f}_jdN_t$ we get that
$d\|\psi\|^2\:=\:0$ and $dP_t\:=\:(aP_ta-P_t)dN_t$. Taking the average we
obtain $\dot{\rho}_t=\:a\rho_ta\:-\:\rho_t$, since $N_t\:-\:\lambda t$ is
a martingale. The above stochastic differential equation admits the
following solution $$\tra\;=\;z_0^1\jp\;+\;z_0^2\jm$$
$$\td\;=\;z_0^1\jm\;+\;z_0^2\jp$$ It implies that
$$P_t\;=\;x_0\jp\;+\;y_0\jm ,$$
where $x_0\:=\:|\psi_0><\psi_0|$ and $y_0\:=\:|\phi_0><\phi_0|,\quad \phi_0
\:=\:(a+a^{\ast})\psi_0\:=\:(z_0^2,\,z_0^1)\prime$.\\

If we take $$a=\left(\matrix{0&1\cr 0&0\cr}\right),$$ as it is usual
in quantum optics problems, then
we have $$\lambda(\psi_t)\;=\;\frac{|\td|^2}{\|\psi_t\|^2}.$$
So we need a point process whose rate function is random and
the situation is slightly more complicated. We have to use the more general
method described at the beginning of this paragraph.

Let us start with calculating
functions $g_i$, which are responsible for the deterministic flow. They are
obtained by taking the derivative of
$$\psi_s\;=\;\frac{exp(-\frac{1}{2}sa^{\ast}a)\psi_t}{\|exp(-\frac{1}{2}s
a^{\ast}a)\psi_t\|}\|\psi_t\|$$
with respect to $s$ and in the instant $s\:=\:0$. So we get
$$g(\psi_t)\;=\frac{1}{2}(-a^{\ast}a\;+<a^{\ast}a>_t)\psi_t.$$
It can be checked that the only functions $f_i$ which lead to the Lindblad
equation are of the following type:
$$f_1(\psi_t)\:=\:-\tra\:+\:\root\of{<\psi_t|\psi_t>}\,e^{ih(\psi_t)},
\quad f_2(\psi_t)\:=\:-\td,$$
where $h:{\bf C}^2\to {\bf R}$ is an arbitrary Lipschitz function. Let us
point out that if we put $e^{ih}\:=\:\td /|\td|$ then we can write $f$ in
a compact form $$f(\psi_t)\;=\;
(\frac{a}{\root\of{<a^{\ast}a>_t}}\;-\;{\bf 1})\psi_t$$
see \cite{barbel91}, but it needs a careful interpretation because zero can
appear in the denominator. Again by simple
calculations we get that $d\|\psi_t\|^2\:=\:0$ and
$$dP_t\;=\left(\matrix{|\td |^2
&-\tra \tdb \cr -\trb \td &-|\td |^2\cr}\right)_{-}
d{\tilde N}_t$$
$$+\frac{1}{2<\psi_t|\psi_t>}\left(\matrix{2|\tra|^2|\td|^2&
\tra\tdb(|\td|^2-|\tra|^2)\cr \trb\td(|\td|^2-|\tra|^2)&-2|\tra|^2|\td|^2
\cr}\right) dt.$$ But $d{\tilde N}_t\:=\:dM_t\:+\:\lambda(\psi_t)dt$ so
after averaging we get the quantum evolution equation for
$\rho_t\:=\:E[P_t]$.

\section{Global existence and uniqueness}

After analyzing a typical example of non uniqueness in the pure quantum
case, here we will return to the general scheme as described in Section 2.
Let $T_t$ be a norm-continuous dynamical semigroup on states of the total
algebra $\at$ corresponding to eq. (\ref{eq:liour}). We extend $T_t$ by
linearity to
the whole predual space $\ats$, which is equal to $\at$, because the total
algebra is finite dimensional. Let $E$ denote a space of all one-dimensional
projectors in $\at$. Because $\at=\oplus_{\alpha = 1}^{\alpha = m}\,
 M(n_{\alpha}\times n_{\alpha})$ we obtain that $E={\dot{\cup}}_{\alpha}
 {\bf C}P_{\alpha}$ and so $E$ is a disjoint sum of compact differentiable
manifolds (complex projective spaces in $\calh_\alpha$). We would like to
associate
with $T_{t}$ a homogeneous Markov -- Feller process with values in $E$ such
that for every $x\in E$
\be T_t (P_x)=\int_E P(t,x,dy)P_y, \label{eq:assoc} \ee
where $P(t,x,dy)$ is the transition probability function for the
process $\xi_t$ and $y\rightarrow P_y$ is a map which assigns to every point
$y\in E$ a one-dimensional projector $P_y$. This leads us to the following
definition.\\
Let $\me$ denote a Banach space of all complex, finite,
Borel measures on $E$. We say that a positive and contractive semigroup
$U_t:\; \me\rightarrow\me$ with a Feller transition function $P(t,x,\Gamma)$
is associated with $T_t$ iff Eq.\ref{eq:assoc} is satisfied.\\
Let us describe this notion more precisely. Let $\pi$ be a map between
two Banach spaces $\me$ and $\ats$ given by
\[  \pi(\mu )=\int_E \, \mu (dx) \, P_x  \]
It is clear that $\pi$ is linear, surjective, preserves positive cones and
$\| \pi \|=1$.

\vspace{5mm}

\noindent{\bf Proposition 1.} $U_t$ is associated with $T_t$ iff $ker\, \pi$
is
$U_t$ -- invariant and $\hut=T_t$, where $\hut$ is the quotient group of
$U_t$ by $ker \, \pi$.

\vspace{5mm}

\noindent{\bf Proof.} Let $U_t$ be associated with $T_t$. It implies that
\[   \int_E \,P(t,x,dy)\, P_y \, =T_t(P_x)  \]
thus for any $\mu_0 \in ker \,\pi$ we have
\[  \int_E (U_t \mu_0 )(dx)\, P_x=\int_E\int_E P(t,y,dx)\,
    \mu_0 (dy)\, P_x= \]
\[  \int_E T_t (P_y)\,\mu_0 (dy)=T_t \, [ \int_E \mu_0 (dy) \, P_y ]
    =0   \]
and so $U_t\mu_0\:\in ker\, \pi$. Moreover $\forall \mu\in \me$
\[  \hut\, \pi(\mu)=\:\pi(U_t \mu )=\int_E (U_t\mu)(dy)\, P_y=  \]
\[  \int_E\int_E \, P(t,x,dy)\mu(dx)\, P_y =T_t[\int_E \mu (dx)\, P_x ]
    =T_t \pi(\mu).  \]
Now let us assume that $\hut =T_t$ i.e. $\forall \mu\in\me$ we have $\hut
\pi(\mu)= T_t \pi(\mu).$ Let us take $\mu =\delta_x$. Then
\[  \hut \pi(\delta_x )=\:\pi(U_t \delta_x )=\int_E (U_t \delta_x)(dy)P_y=
\]
\[  \int_E\int_E P(t,z,dy)\delta_x (dz) P_y =\int_E P(t,x,dy) P_y  \]
and $T_t \pi(\mu )=T_t(P_x)$ so $T_t (P_x)=\int_E P(t,x,dy)P_y.$
$\Box $

\vspace{5mm}

It means that to find $U_t$ is to extend the semigroup $T_t$ from $\me /
ker\, \pi$
to $\me$ in an invariant way. It should be emphasized that, in general, such
an `extension' may not exist or, if it exists, need not be unique. We show
that
in our case, under mild assumptions, the existence and the uniqueness can be
proved.

Let us write the evolution equation for states in the Lindblad form
\[  \dot{\rho}=-i [ H,\rho ]+\sum_k V_k^{\, *}\rho V_k -\frac{1}{2}
    \{ \rho , \sum_k V_k V_k^{\, *} \},  \]
where $H=diag(H_1 , \ldots ,H_m),$ $\ha =H^{\ast}_{\alpha}
\in M(n_\alpha\times n_\alpha )$
and $V_k$ satisfy the following assumptions:

\ \ \  a) $(V_k)_{\alpha\alpha} =0 $ for every $k$ and $\alpha$

\ \ \  b) if for some $k,l,\alpha,\beta$ $(V_k)_{\alpha\beta}\neq 0$ and
          $(V_l)_{\alpha\beta}\neq 0$ then $k=l$\footnote{
                In general we can allow for a weaker version:
                $(V_k)_{\alpha\beta}\neq 0$ and $(V_l)_{\alpha\beta}\neq 0$
                $\Rightarrow \exists c\in {\bf C}\, :\,
                (V_k)_{\alpha\beta}=c(V_l)_{\alpha\beta},$
                but this simply reduces to b) above by substitution
            $(\tilde{V}_k)_{\alpha\beta}:=\sqrt{1+|c|^2} (V_k)_{\alpha\beta}$
                and
                $(\tilde{V}_l)_{\alpha\beta}=0$ for $k\neq l.$    }

\vspace{5mm}

Let $A$ be a densely defined linear operator on $C(E)$ with $D(A)=C^1(E)$
given by
\[  (Af)(x)=\sum_{\alpha\neq\alpha_0} c_\alpha (x)f(x_\alpha)-c(x)f(x)+v(x)f,
\]
where $x\in {\bf C}P_{\alpha_0},$ \ $c_\alpha (x)=\tr (P_x W_{\alpha_0
\alpha}
W_{\alpha_0 \alpha}^{\, *}),$ \ $W_{\alpha_0 \alpha}=\sum_k (V_k)_{\alpha_0
\alpha}\,\in L(\ha,{\cal H}_{\alpha_0}),$ \
$W_{\alpha_0 \alpha}^{\, *}=\sum_k (V_k)_{\alpha_0 \alpha}^{\, *}
\,\in L({\cal H}_{\alpha_0},\ha),$ \
$c(x)=\sum_{\alpha\neq\alpha_0} c_\alpha (x),$ \
$P_{x_\alpha}=\frac{W_{\alpha_0\alpha}^{\, *}P_x W_{\alpha_0\alpha}}
{\tr (P_x W_{\alpha 0\alpha}W_{\alpha_0\alpha}^{\, *})} \in {\bf C}P_\alpha$
and $x\rightarrow v(x)$ is a vector field on $E$ such that
\[  v(x)=-i [H_{\alpha_0},P_x ]-\frac{1}{2}\{ P_x, \sum_{\alpha\neq\alpha_0}
\waoa\waoa^{\,*} \}+P_x \tr (P_x \sum_{\alpha\neq\alpha_0}\waoa\waoa^{\, *})
\]
It may be easily checked that $v(x)\in T_x {\bf C}P_\alpha =\:T_x E$.
Because
\[  g_t(P_x)=\frac
    {\exp[t(-iH_{\alpha_0}-\frac{1}{2}\waoas )] P_x
     \exp [t(iH_{\alpha_0}-\half \waoas)]}
    {\tr (P_x \exp[-t\waoas])}    \]
is an integral curve for $v$, so we have that $v$ is a complete vector field.

\vspace{5mm}

\noindent{\bf Theorem 2.} $A$ is a generator of a strongly continuous
positive semigroup
of contractions $S_t$ on $C(E)$.

\vspace{5mm}

\noindent{\bf Proof.} $A=A_1+A_2$, where $(A_1
f)(x)=\sum_{\alpha\neq\alpha_0}
c_\alpha (x)\delta_{x_\alpha}f-c(x)\delta_x f$ and $A_2=v.$
It is clear that $A_1$ is a bounded and dissipative operator. It is also
a dissipation i.e. $A_1(f^2)\geq 2fA_1(f)$ for $f={\bar f}$.
Because $A_2$ generates a flow on $E$ given by $f(x)\rightarrow f(g_t (x)),$
where $g_t (x)$ is the integral curve of $v$ starting at the point $x$, it
follows that $A=A_1+A_2$ is the generator of a strongly continuous
semigroup of contractions (see for example \cite{goldstein}).
Positivity follows from the Trotter
product formula, since both $A_1$ and $A_2$ generates positive
semigroups. $\Box$

Let $P(t,x,\Gamma)$ denote the transition function of $S_t$.

\vspace{5mm}

\noindent{\bf Proposition 3.} $P(t,x,\Gamma)$ is a Feller transition
function.

\vspace{5mm}

\noindent{\bf Proof.} By theorem 2.8 of \cite{dynkin}, and by
conservativeness
of $P(t,x,\Gamma)$, it is enough to show that for every $x\in E$ and for any
$f \in C^1(E)$ such that $f(x)=0,$ $f(y)\leq 0$ $\forall y\in E$ we have
$(Af)(x)\leq 0.$ Because $f$ has a maximum at $x$ so $(A_2 f)(x)=0.$
Moreover, as $x\in\cpao$ for some $\alpha_0$ and $c_\alpha (x)\geq 0$
$\forall \alpha\neq\alpha_0$ we have

\newpage

\[  (A_1f)(x)=\sum_{\alpha\neq\alpha_0} c_\alpha (x)f(x_\alpha)\leq 0 \]
$\Box$

\vspace{5mm}
Now prove that that our process reproduces $T_t$.

\noindent{\bf Theorem 4.}  Let $(U_t\mu)(\Gamma):=\int_E
P(t,x,\Gamma)\mu(dx)$ for
$\mu\in\me.$ Then $U_t$ is associated with $T_t$.

\vspace{5mm}

\noindent{\bf Proof.} At first we show that $\forall x\in E$
\be  L(P_x )=[A(P)](x),  \label{thesis}   \ee
where $L$ is the generator of $T_t$, $A$ is the generator of $S_t$ and
$P\, : \, x\rightarrow P_x.$

Let $x\in\cpao$. In $\calh= \oplus_{\alpha=1}^{m}\calh_\alpha$ let us choose
any orthonormal basis $\{ e_{\alpha,i_\alpha}\}^{\alpha=1,\ldots ,m}_
{i_\alpha =1,\ldots ,n_\alpha },$ for which $e_{\alpha,i_\alpha}\in
\calh_{\alpha}$. Obviously, for any
$P_x\in\ats$ \\ $< e_{\alpha,i_\alpha}|L(\rho)|
e_{\beta,i_\beta} > \: =0$ for $\alpha\neq\beta$ and the same is true for
$[A(P)](x)$. So it is enough to
evaluate the $(\beta,i_\beta,j_\beta)$-th matrix elements of both
sides of Eq.(\ref{thesis}):
\[  <\ebib|[A(P)](x)|\ebjb >\: =\sum_{\alpha\neq\alpha_0}\tr (P_x\waoa
    \waoa^{\, *})\cdot\frac{<\ebib|\waoa^{\, *}P_x\waoa |\ebjb >}
    {\tr (P_x \waoa\waoa^{\, *}) }-  \]
\[  - \sum_{\alpha\neq\alpha_0}\tr (P_x\waoa\waoa^{\, *} )
                                      <\ebib |P_x |\ebjb > +  \]
\[ <\ebib |(-i[H_{\alpha_0},P_x ]-\half \{P_x,\waoas\}+P_x\,\tr
   (P_x \waoas ))|\ebjb >\: =   \]
$$  = \, <\ebib |\waob^{\, *} P_x \waob |\ebjb >\: + \delta_{\alpha_0\beta}
     <\ebib |(-i[H_{\alpha_0},P_x ]-$$
\be     \half\{ P_x,\waoas \} ) | \ebjb >
                                 \label{merhs}   \ee
On the other hand the $\beta$-th component of $L(P_x)$
\[  (L(P_x))_\beta =\sum_k (V_k)_{\alpha_0\beta}^{\,
*}P_x(V_k)_{\alpha_0\beta}
    +\delta_{\alpha_0\beta} -(i[H_{\alpha_0},P_x ]+\half\{ P_x,
\sum_{k,\alpha}
     (V_k)_{\beta\alpha}(V_k)_{\beta\alpha}^{\, *} \} )=   \]
\be  =\waob^{\, *}P_x\waob +\delta_{\alpha_0\beta}( -i [H_{\alpha_0},P_x ]
     -\half \{ P_x, \waoas \} )  \label{lhs}   \ee
Where the last equality holds owing to assumptions a) and b) above. Taking
the $(\beta,i_\beta,j_\beta)$-th matrix element of (\ref{lhs}) we see that
it coincides with (\ref{merhs}), thus, due to arbitrariness of
$(\beta,i_\beta,
j_\beta )$, we have proved Eq. (\ref{thesis}). \\
Let $F$ denote the finite dimensional space of functions generated by
$x\to <\psi|P_x|\phi>$. It is clear that $F\:=\:\{f:f(x)=Tr(AP_x),\:A\in
{\cal A}_T\}$. So $dim\,F\:=\:dim\,{\cal A}_T$. We show that $F$ is the null
space for $ker\,\pi$. Let $f(x)\:=\sum_{i,j}<\psi_i|P_x|\psi_j>$ and let
$\mu_0\in ker\,\pi$. Then
$$\mu_0(f)\;=\int \mu_0(dx)f(x)\;=\sum\limits_{i,j}<\psi_i|\int \mu_0
(dx)P_x|\psi_j>\;=\;0$$
Moreover because $(A<\psi_i|P|\psi_j>)(x)\:=\:<\psi_i|L(P_x)|\psi_j>$ we
have that $A:F\to F$ and so $S_t:F\to F$. It implies that
$U_t:ker\,\pi\to ker\,\pi$
since $U_t\mu(f)\:=\:\mu(S_tf)$.
Let $\hut$ be the quotient semigroup. Then
$$\lim\limits_{t\to 0}\frac{1}{t}[\hut(P_x)\:-\:P_x]\;=
\lim\limits_{t\to 0}\frac{1}{t}[\pi(U_t\delta_x)\:-\:P_x]\;=$$
$$\lim\limits_{t\to 0}(\int\int P(t,\,z,\,dy)\delta_x(dz)P_y\:-\:P_x)\;=
\;(AP)(x),$$ so $\hut$ and $T_t$ have the same generator and thus coincide.
By Prop. 1 $U_t$ is associated with $T_t$. $\Box$

\vspace{5mm}

We can pass to the uniqueness problem.
Let us consider a class of Markov processes associated with a general
nonsymmetric Dirichlet form ${\cal E}$ on $L^2 (E,dm)$ (here
$dm|_{CP_\alpha}$
is a positive $U({\cal H}_\alpha)$-invariant measure on Borel sets
on ${\bf C}P_\alpha$) given by the closure of:
\[  {\cal E}(u,v)=\int_E T(du,dv)\, dm+\int_E u(X.v)\, dm+\int_E (Y.u)v\, dm+
    \int_E uvc\, dm+  \]
\be  +\int_{E\times E\setminus\Delta} (u(x)-u(y))\, (v(x)-v(y))\, J(dx,dy)
     \label{form}  \ee
for $u,v\in C^\infty (E)$. In (\ref{form}) we have :
\begin{itemize}
  \begin{item}
        $X,Y$ -- smooth vector fields on $E$  \end{item}
  \begin{item}
        $c\in C^\infty (E)$  \end{item}
  \begin{item}
        $T$ -- smooth (2,0)-tensor field, positively defined: $T(du,du)\geq
0$
        for any $u\in C^\infty (E)$  \end{item}
  \begin{item}
        $\Delta=\{ (x,x)\in E\times E\}$ -- the diagonal  \end{item}
  \begin{item}
        $J(dx,dy)$ -- positive symmetric Radon measure on $E\times E
        \setminus\Delta$ satisfying:
        \begin{itemize}
              \begin{item}
                    $\int_{E\times E\setminus\Delta}(u(x)-u(y))^2\,J(dx,dy)
                    <\infty$ for any $u\in C^\infty (E)$  \end{item}
              \begin{item}
                    the Radon derivative $\frac{J(dx,dy)}{dm(x)}$ exists and
is
                    a Borel measure  \end{item}
         \end{itemize}
         \end{item}
  \begin{item}
        for any $u\geq 0$ hold: $\int_E (cu+X.u)\, dm ,\: \int_E (cu+Y.u)\,
        dm \geq 0$   \end{item}
\end{itemize}
It is worth to emphasize that such Dirichlet forms contain jumps,
deterministic flows and diffusion processes as well. A straightforward
calculation leads to the following result:
\vspace{5mm}

\noindent{\bf Theorem 5.} The generator $B$ of the Dirichlet form
${\cal E}$ defined by
(\ref{form}) is given, in some coordinate system, by the formula
\be (Bu)(x)=(fu)(x)+\sum_i V^i(x)(\partial_i u)(x)
+\sum_{ij}T^{ij}(x)(\partial_i\partial_j u)(x)-\int_E\mu (x,dy)\, u(y)
    \label{formgen}  \ee
where $f\in C^\infty (E),$ \ $V$ -- smooth vector field on $E,$ \ $T$ --
smooth,
positive (2,0)-tensor field and $\mu$ -- family of Borel signed measures,
$u\in
C^\infty (E)$ (the domain of $B$ comes from closing $(B,C^\infty (E))$ ).
The detailed form of $f,$ $V$ and $\mu$ is given by:
\[  f=\frac{1}{M}X.M+\partial_i X^i -c  \]
\[  V=\frac{1}{M}T^{ij}(\partial_j M)\partial_i+(\partial_j T^{ij}
)\partial_i
    +X-Y   \]
\[  \mu (x,dy)= 2 [ \delta_x (dy)\frac{\int_{E\setminus \{ x\} }J(dx,
    \stackrel{\downarrow}{dz} )}{dm(x)}-\frac{J(dx,dy)}{dm(x)} ]  \]
and
\[  \mu (x,\{ x\})= 2 [ \frac{\int_{E\setminus \{ x\} }J(dx,
    \stackrel{\downarrow}{dz} )}{dm(x)}], \]
where we have used the following notation
\begin{itemize}
  \begin{item}
        $M$ is a coordinate of the volume form: $dm(x)=M(x)dx^1\wedge\ldots
        \wedge dx^{2k}$ for $x\in{\bf C}P^k $     \end{item}
  \begin{item}
        $\delta_x (\cdot)$ -- a measure concentrated in $\{ x\}$  \end{item}
  \begin{item}
        an arrow indicates the variable the integral is evaluated over
        \end{item}
\end{itemize}

\vspace{5mm}

\noindent{\bf Remark 1.} The generator $B$ may be written in a fully
invariant way:

\[  (Bu)(x)=-cu-Y.u+\frac{1}{dm}\, L_{[T(du)+uX]}\, dm  \]
where $L_Z\,\omega$ means the Lie derivative of the form $\omega$, associated
to the vector field $Z$.

\vspace{5mm}

\noindent{\bf Remark 2.} The proof of the Theorem 5 is straightforward -- one
should
use the Stokes theorem for compact oriented manifold for the form
\[  \alpha =v\,i_{[T(du)+uX]}\, dm  \]
and because $E$ is without boundary ($\partial E =\emptyset $) we have
$\int_E d\alpha = 0$. Evaluating $d\alpha$ we obtain the form of $B$.
\vspace{5mm}

Using Theorem 5 and the property that $\tr[L(P_x)]\:=\:0\;\forall x\in E$
we conclude that $B({\bf 1})\equiv 0$, {\bf 1}-denotes the constant
function taking value 1, and so $B$ can be written in the following
form:
$$ (Bu)(x)\:=\sum_{ij} T^{ij}(x)(\partial_i\partial_j u)(x)\:+
\sum_i V^i(x)(\partial_i u)(x)\:+$$
\be \int_E {\mu}_0(x,dy)u(y)\;-\;{\mu}_0(x,E)u(x), \ee
where $(T^{ij}(x))$ form a positive matrix and ${\mu}_0(x,dy)$ is a
positive measure such that ${\mu}_0(x,\{x\})\:=\:0$ for every $x\in E$.
Its domain $D(B)$ consists of $C^2$-functions.

\vspace{5mm}

\noindent{\bf Lemma 6.}
$(V_k)_{\alpha\alpha}\:=\:0\Rightarrow\forall\alpha\in\{1,
\ldots ,m\}\;\forall x,y\in {\bf C}P_{\alpha}$ such that $P_x\bot P_y$
the equality $Tr[P_yL(P_x)]\:=\:0$ is satisfied.\\
{\bf Proof}. let $x,y\in {\bf C}P_{\alpha}$ and $P_x\bot P_y$. Then
$$Tr[P_yL(P_x)]\:=\:-iTr(P_y[H_{\alpha},\;P_x])\:+\sum_k Tr[P_y(V_k^{\ast}
P_xV_k)_{\alpha\alpha}]\:-$$
$$\frac{1}{2}\sum_k Tr[P_y\{P_x,\;(V_kV_k^{\ast})_{\alpha\alpha}\}]\:=\:
\sum_k Tr[P_y(V_k^{\ast}P_xV_k)_{\alpha\alpha}]$$
But $$(V_k^{\ast}P_xV_k)_{\alpha\alpha}=\:(V_k)^{\ast}_{\alpha\alpha}P_x
(V_k)_{\alpha\alpha}\:=\:0$$
so the assertion follows. $\Box$\\

\vspace{5mm}
We are now in position to show that the diffusion part is necessarily
zero.
\vspace{5mm}
\noindent{\bf Theorem 7.} $T^{ij}(x)\equiv 0$ for every i, j.\\
\noindent{\bf Proof.} Because
$$B[Tr(P_yP)](x)\:=\:Tr[P_yL(P_x)]$$
so, by the above lemma, for every $\alpha$ and every $x,y\in {\cpa}$
such that $P_y\bot P_x$
we have that $B[Tr(P_yP)](x)\:=\:0$. Let us denote the function
$z\to Tr(P_yP_z)$ by $f_y(z)$. Then
$$(Bf_y)(x)\:=\int_{\cpa} {\mu}_0(x,dz)f_y(z)\;+\sum_{ij}T^{ij}(x)
(\partial_i\partial_jf_y)(x)\;+\sum_i V^i(x)(\partial_if_y)(x)$$
It is clear that $f_y$ is a smooth function and possesses a minimum
at point $x$. So $\sum_i V^i(x)(\partial_if_y)(x)\:=\:0$ and we arrive at
$$\int_{\cpa} {\mu}_0(x, dz)f_y(z)\;+\sum_{ij}T^{ij}(x)(\partial_i
\partial_jf_y)(x)\:=\:0$$
But $(\partial_i\partial_jf_y(x))$ and $(T^{ij}(x))$ are positive
matrices so, by Schur's lemma, $(T^{ij}(x)\partial_i\partial_jf_y(x))$
is also a positive matrix. It follows that $$\sum_{ij} T^{ij}(x)
\partial_i\partial_jf_y(x)\:=\:0$$
Now let us introduce a chart at point $x$, let say, $x$ = [(1,0,...,0)],
$(U_0,\phi_0)$ such that
$$U_0\:=\:\{[(z_0,z_1,\dots ,z_{n-1})]:\;z_i\in {\bf C},\;\sum_i
|z_i|^2=\:1,\;z_0\neq 0\}$$
$$\phi_0[(z_0,z_1,\dots ,z_{n-1})]\:=(\frac{z_1}{z_0},\dots
,\frac{z_{n-1}}{z_0})\:=\:(x_1,y_1,\dots ,x_{n-1},y_{n-1}),$$
where $x_i\:=\:Re\frac{z_i}{z_0}$, $y_i\:=\:Im\frac{z_i}{z_0}$. Then
$\phi_0(x)\:=\:\vec{0}\in {\bf R}^{2(n-1)}$. Let us choose
$y$ = [(0,1,0,...,0)]. It is clear that $P_y\bot P_x$ and so
$$\sum_{i,j=1}^{n-1} [T^{ij}_{x,x}(x)\frac{\partial^2(f_y\circ\phi_0^{-1})}
{\partial x_i\partial x_j}(\vec{0})\:+\:2T^{ij}_{x,y}(x)\frac{\partial
^2(f_y\circ\phi_0^{-1})}{\partial x_i\partial y_j}(\vec{0})\:+$$
$$T^{ij}_{y,y}(x)\frac{\partial^2(f_y\circ\phi_0^{-1})}{\partial
y_i\partial y_j}(\vec{0})]\;=\:0$$
But for every $j\geq 2$ we have
$$\frac{\partial^2(f_y\circ\phi_0^{-1})}{\partial x_j^2}(\vec{0})\:=$$
$$\lim_{h\to{\infty}} \frac{1}{h}[\frac{\partial(f_y\circ\phi_0^{-1})}
{\partial x_j}(0,\dots ,x_j=h,0,\dots ,0)\:-\:\frac{\partial(f_y\circ
\phi_0^{-1})}{\partial x_j}(\vec{0})]\:=\:0$$
In the same way we prove that for every $j\geq 2$
$$\frac{\partial^2(f_y\circ\phi_0^{-1})}{\partial y_j^2}(\vec{0})\:=\:0$$
By positivity of the matrix $D^2(f_y\circ\phi_0^{-1})(\vec{0})$ we obtain
that $$T^{11}_{x,x}(x)\frac{\partial^2(f_y\circ\phi_0^{-1})}{\partial
x_1^2}(\vec{0})\:+\:2T^{11}_{x,y}(x)\frac{\partial^2(f_y\circ\phi_0^{-1})}
{\partial x_1\partial y_1}(\vec{0})\:+\:T^{11}_{y,y}(x)\frac{\partial^2
(f_y\circ\phi_0^{-1})}{\partial y_1^2}(\vec{0})\:=\:0$$
Let $\lambda$ be an embedding $\lambda :{\bf C}P^1\to{\cpa}$ given by
$$\lambda [(z_0,z_1)]\:=\:[(z_0,z_1,0,\dots ,0)]$$
It is clear that $x\:=\:\lambda (\vec{n_0})$ and $y\:=\:\lambda (\vec{n})$
for some unique $\vec{n_0},\:\vec{n}\in {\bf C}P^1\:=\:S^2$. Let $\psi_0$
be a chart at $\vec{n_0}$ given by
$$\psi_0:\;{\bf C}P^1-\{\vec{n}\}\to{\bf C},\quad\psi_0(\vec{m})\:=\:
p\circ\phi_0\circ\lambda (\vec{m}),$$
where $p\:=\:{\bf C}^n\to{\bf C}$ is the projection onto the first
coordinate. So we may write that
$$a^{11}(\vec{n_0})\frac{\partial^2(f_{\vec{n}}\circ\psi_0^{-1})}{\partial
q_1^2}(\vec{0})\:+\:2a^{12}(\vec{n_0})\frac{\partial^2(f_{\vec{n}}\circ
\psi_0^{-1})}{\partial q_1\partial q_2}(\vec{0})\:+$$ $$+a^{22}(\vec{n_0})
\frac{\partial^2(f_{\vec{n}}\circ\psi_0^{-1})}{\partial q_2^2}(\vec{0})
\:=\:0,$$
where $a^{11}(\vec{n_0})\:=\:T^{11}_{x,x}(x)$, $a^{12}(\vec{n_0})\:=\:
T^{11}_{x,y}(x)$, $a^{22}(\vec{n_0})\:=\:T^{11}_{y,y}(x)$ and $q_1(
\vec{m})\:=\:x_1(\lambda(\vec{m}))$, $q_2(\vec{m})\:=\:y_1(\lambda
(\vec{m}))$. Let us change the chart $\psi _0$ onto spherical
coordinates $(\theta,\varphi)$, $0\leq\theta\leq\pi$, $0\leq\varphi
\leq 2\pi$ in such a way that $\theta(\vec{n_0})\:=\:\pi/2$, $\varphi
(\vec{n_0})\:=\:0$ i.e. $\vec{n_0}\:=\:(1,0,0)$ and $ \theta(\vec{n})\:
=\:\pi/2$, $\varphi(\vec{n})\:=\:\pi$ i.e. $\vec{n}\:=\:(-1,0,0)$.
Because $$f_{\vec{n}}(\vec{m})\:=\:Tr(P_{\vec{n}}P_{\vec{m}})\:=\:
\frac{1}{2}(1\:+\:<\vec{n},\vec{m}>)\:=\:\frac{1}{2}(1\:-\:\sin\theta
\cos\varphi)$$ so
$$\frac{\partial^2f_{\vec{n}}}{\partial\theta\partial\varphi}(\vec{n_0})
\:=\:0,\quad \frac{\partial^2f_{\vec{n}}}{\partial\theta^2}(\vec{n_0})\:
=\:\frac{\partial^2f_{\vec{n}}}{\partial\varphi^2}(\vec{n_0})\:=\:
\frac{1}{2}$$
which implies that $\tilde{a}^{11}(\vec{n_0})\:=\:\tilde{a}^{12}(\vec{n_0})
\:=\:\tilde{a}^{22}(\vec{n_0})\:=\:0$, where $\tilde{a}^{ij}$ are the
coefficients in the chart $(\theta,\varphi)$. But it is equivalent to
$$T^{11}_{x,x}(x)\;=\;T^{11}_{x,y}(x)\;=\;T^{11}_{y,y}(x)\;=\;0$$
Changing $y\:=\:[(0,1,0,\dots ,0)]$ into $y\:=\:[(0,0,1,0,\dots ,0)]$
we obtain that $$ T^{22}_{x,x}(x)\;=\;T^{22}_{x,y}(x)\;=\;T^{22}_{y,y}
(x)\;=\;0$$ and so on. Thus, by the positivity, $T^{ij}(x)\:=\:0$ for
every $j,k$. Because $x$ was arbitrary the assertion follows. $\Box$\\

{}From the above theorem we conclude that the generator $B$ has to be
of the following form
$$Bu(x)\:=\:V(x)u\:+\int_E \mu_0(x,dy)u(y)\:-\:\mu_0(x,E)u(x)$$
with domain $D(B)\:=\:C^1(E)$ as $B$ is a closed operator.
To proceed further we first need a lemma.

\vspace{5mm}

\noindent{\bf Lemma 8.} Let $X$ be a tangent vector to ${\cpa}$ at point
$P_x$. Then
$P_x\:+\:X\geq 0\Leftrightarrow X\:=\:0$.\\
{\bf Proof.} Because $X\in T_x{\cpa}$ so $P_xX\:+\:XP_x\:=\:X$. It
implies that $P_xXP_x\:=\:0$ and $P_x^{\perp}XP_x^{\perp}\:=\:0$, where
$P_x^{\perp}\:=\:I-P_x$. It means that in a basis $P_x{\cal H}\oplus P_x
^{\perp}{\cal H}\quad X$ is of the form
$\left( \begin{array}{cc}
0&X^{\ast}\\ X&0 \end{array} \right)$.
So $P_x\:+\:X$ is a positive matrix if and only if $X\:=\:0$. $\Box$

\vspace{5mm}

\noindent{\bf Theorem 9.} $B\:=\:A$.\\
{\bf Proof.} Because $A$ and $B$ are generators of semigroups which are
associated with $T_t$ so for every $x\in E$ we have that $[(B-A)P](x)\:
=\:0$. Let $x\in {\cpao}$. Then
$$V(x)P\:+\sum_{\alpha =1}^m\int_{\cpa} \mu_{0,\alpha}(x,dy)P_y\:-\:
\mu_0(x,E)P_x\:-$$ $$\sum_{\alpha\neq\alpha_0} c_{\alpha}(x)P_{x_{\alpha}}
\:+\:c(x)P_x\:-\:v(x)P\:=\:0,$$ where $\mu_{0,\alpha}(x,dy)$ denotes the
restriction of $\mu_0(x,dy)$ onto ${\cpa}$.
It is an operator valued equation so it has to be satisfied for every
$\alpha$ separately. So for any $\alpha\neq {\alpha}_0$ we get
$$\int_{\cpa} \mu_{0,\alpha}(x,dy)P_y\;=\;c_{\alpha}(x)P_{x_{\alpha}}$$
which implies that $\mu_{0,\alpha}(x,dy)\:=\:c_{\alpha}(x)\delta(x_
{\alpha})(dy)$. For $\alpha_0$ we have
$$\int_{\cpao} \mu_{0,\alpha_0}(x,dy)P_y\:-\:\mu_0(x,E)P_x\:+\:c(x)P_x\:
+V(x)\:-\:v(x)\:=\:0$$
Let us introduce $a(x)\:=\:c(x)\:-\:\mu_0(x,E)$ and $w(x)\:=\:V(x)\:-\:
v(x)$. Then taking the trace of the above equation we obtain $a(x)\leq 0$.
Let us assume that $a(x)<0$. It implies that
$$\frac{1}{|a(x)|}\int_{\cpao} \mu_{0,\alpha_0}(x,dy)P_y\:=\:P_x\:-\:
\frac{1}{|a(x)|}w(x)$$
The left hand side of the above equation gives a positive operator
and $w(x)\in T_x{\cpao}$ so,
by Lemma 8, $w(x)\:=\:0$. Thus we arrive at the contradiction
because $\mu_{0,\alpha_0}
(x,\{x\})\:=\:0$. So $a(x)\:=\:0$ and we obtain that
$$\int_{{\bf C}P_{\alpha 0}} \mu_{0,\alpha 0}(x,\,dy)P_y\:+\:w(x)\:=\:0$$
Evaluating the trace we get that $\mu_{0,\alpha_0}
(x,{\cpao})\:=\:0$. Because it is a positive measure so it vanishes
on every Borel subset of ${\cpao}$. So $w(x)\:=\:0$ too and hence
$A\:=\:B$. $\Box$

\vspace{5mm}

Thus we have the uniqueness. In the proof above we used
repeatedly the fact that our Hilbert spaces were finite dimensional.
In an infinite dimensional case the problem is much harder and
we have no rigorous result. Our intuition is shaped here only by the
infinitesimal argument of Section 3.
\section{Conclusions}
We have seen that the special class of couplings between
a classical and a quantum system leads to a unique piecewise
deterministic process on pure states of the total system that
after averaging recovers the original master Liouville
equation for statistical states. Irreversibility of the
master equation describing time evolution of ensembles is reflected
by going from potential to actual in the course of quantum jumps
that accompany classical events. That is all fine but a natural
question arises: what is classical? There are several options
possible when answering this question. First of all the theory
may be considered as phenomenological - then we choose as
classical this part of the measurement apparatus (or observer)
whose quantum nature is simply irrelevant for the given problem.
Second, we may think of superselection quantities \cite{amman,lands94}
as truly classical variables. Some of them may play an important role
in the dynamics of the measurement process - this remains for a while
just a hypothesis. It is to be noticed that Jibu et. al (cf. \cite{jibu94},
especially the last section \lq Quantum Measurement by Quantum Brain\rq
 puts forward a similar hypothesis in relation to the possible role of
microtubules in the quantum dynamics of consciousness.\\
Finally, a careful reader certainly noticed that in the formalism
of EEQT one never really needs $C$ to be a {\sl classical}\, system.
It can be a quantum system as well. What is important it is that the
the Liouville evolution preservers the diagonal of $C$. Thus the end product
of the decoherence program \cite{decoa,decob,decoc},
can be directly fed into the EEQT event engine. The uniqueness result above -
will be immediately relevant also for this case.\\

\vskip10pt
\noindent
{\bf Acknowledgements}\\
The early version of this paper was written when one of us (A.J)
was visiting RIMS, Kyoto U. Thanks are due to Prof. H. Araki and
for his kind hospitality and to Japanese Ministry of Education
for the extended support. The third named author (R.O) acknowledges
support of the Polish KBN grant no 2P30205707. A.J. also thanks for
the support of the A. von Humboldt Foundation. We owe to
Prof. Ph. Blanchard many discussions, encouragement and hospitality
at BiBoS.

\end{document}